\documentclass[aps,prl,floatfix,superscriptaddress,twocolumn]{revtex4}

\usepackage{amsmath}
\usepackage{amssymb}
\usepackage{graphicx}
\usepackage{epstopdf}
\usepackage{color}
\usepackage{cancel}

\newcommand*{\comment}[1]{}
\newcommand*{\tr}{\mathrm{tr}}
\newcommand*{\ket}[1]{| #1 \rangle}
\newcommand*{\bra}[1]{\langle #1 |}

\newcommand*{\ketbra}[2]{|#1\rangle\!\langle#2|}

\newcommand*{\id}{\openone}
\newcommand*{\ot}{\otimes}

\newcommand{\m}{\,\mbox{m}}

\newcommand{\mm}{\,\mbox{mm}}

\newcommand{\nm}{\,\mbox{nm}}

\newcommand{\s}{\,\mbox{s}}

\newcommand{\ns}{\,\mbox{ns}}

\newcommand{\kHz}{\,\mbox{kHz}}

\newcommand{\mW}{\,\mbox{mW}}

\begin{document}

\title{Experimental investigation of the uncertainty principle\\ in the presence of quantum memory}

\author{Robert Prevedel}\email{robert.prevedel@iqc.ca}
 \affiliation{Institute for Quantum Computing, University of Waterloo, Waterloo, N2L 3G1, ON, Canada}
\author{Deny R. Hamel}
 \affiliation{Institute for Quantum Computing, University of Waterloo, Waterloo, N2L 3G1, ON, Canada}
\author{Roger Colbeck}
\affiliation{Perimeter Institute for Theoretical Physics, 31 Caroline Street North, Waterloo, Ontario N2L 2Y5, Canada}
 \author{Kent Fisher}
\affiliation{Institute for Quantum Computing, University of Waterloo, Waterloo, N2L 3G1, ON, Canada}
\author{Kevin J. Resch}
\affiliation{Institute for Quantum Computing, University of Waterloo, Waterloo, N2L 3G1, ON, Canada}

\begin{abstract}
  Heisenberg's uncertainty principle provides a fundamental
  limitation on an observer's ability to simultaneously predict the
  outcome when one of two measurements is performed on a quantum
  system. However, if the observer has access to a particle (stored
  in a quantum memory) which is entangled with the system, his
  uncertainty is generally reduced.  This effect has recently been
  quantified by Berta et al.\ [Nature Physics 6, 659 (2010)] in a
  new, more general uncertainty relation, formulated in terms of
  entropies. Using entangled photon pairs, an optical delay line
  serving as a quantum memory and fast, active feed-forward we
  experimentally probe the validity of this new relation. The
  behaviour we find agrees with the predictions of quantum theory and
  satisfies the new uncertainty relation.  In particular, we find
  lower uncertainties about the measurement outcomes than would
  be possible without the entangled particle. This shows not only
  that the reduction in uncertainty enabled by entanglement can be
  significant in practice, but also demonstrates the use of the
  inequality to witness entanglement.
\end{abstract}

\maketitle

Consider an experiment in which one of two measurements is made on a
quantum system.  In general, it is not possible to predict the
outcomes of both measurements precisely, which leads to uncertainty
relations constraining our ability to do so.  Such relations lie at
the heart of quantum theory and have profound fundamental and
practical consequences.  They set fundamental limits on precision
technologies such as metrology and lithography, and also served as the
intuition behind new types of technologies such as quantum
cryptography~\cite{Wiesner83,Bennett1984}.

The first relation of this kind was formulated by Heisenberg for the
case of position and momentum~\cite{Heisenberg1927}. Subsequent work  by Robertson~\cite{Robertson1929} 
and Schr\"odinger~\cite{Schroedinger1930} generalized this relation to arbitrary pairs of observables.
In particular, Robertson showed that
\begin{equation}\label{Robertson}
    \Delta R\cdot\Delta S \geq \frac{1}{2}\left|\langle\left[R,S\right]\rangle\right|,
\end{equation}
where uncertainty is characterized in terms of the standard deviation
$\Delta R$ for an observable $R$ (and likewise for $S$) and the
right-hand-side (RHS) of the inequality is expressed in terms of the
expectation value of the commutator, $\left[R,S\right]:=RS-SR$, of the
two observables.

More recently, driven by information theory, uncertainty relations
have been developed in which the uncertainty is quantified using
entropy~\cite{Bialynicki-Birula1975,Deutsch1983}, rather than the
standard deviation. This links uncertainty relations more naturally to
classical and quantum information and overcomes some pitfalls of
equation~\eqref{Robertson} pointed out by Deutsch~\cite{Deutsch1983}.
Most uncertainty relations apply only in the case where the
uncertainty is measured for an observer holding only classical
information about the system.  One such relation, conjectured by
Kraus~\cite{Kraus1987} and subsequently proven by Maassen and
Uffink~\cite{Maassen1988}, states that for any observables
$R$ and $S$
\begin{equation}\label{MaassenUffink}
   H(R)+H(S) \geq \log_{2}\frac{1}{c},
\end{equation}
where $H(R)$ denotes the Shannon entropy~\cite{Shannon1949} of the
probability distribution of the outcomes when $R$ is measured and the
term $1/c$ quantifies the \emph{complementarity} of the observables.
For non-degenerate observables, it is defined by $c:=\max_{r,s}|\langle\Psi_r|\Upsilon_s\rangle|^2$, where
$\ket{\Psi_r}$ and $\ket{\Upsilon_s}$ are the eigenvectors of $R$ and $S$,
respectively.

Interestingly, the above relations do not apply to the case of an
observer holding quantum information about the measured system.  In
the extreme case that the observer holds a particle maximally
entangled with the quantum system, he is able to predict the outcome
precisely for both choices of measurement.  This dramatically
illustrates the need for a new uncertainty relation.

Recently, Berta \emph{et al.}~\cite{Berta2010} derived such a relation (an
equivalent form of this relation had previously been conjectured by
Renes and Boileau~\cite{Renes2009}).  The new relation is
\begin{equation}\label{Berta}
   H(R|B)+H(S|B) \geq \log_{2}\frac{1}{c}+H(A|B),
\end{equation}
where the measurement ($R$ or $S$) is performed on a system, $A$, and
the additional quantum information held by the observer is in $B$.
The Shannon entropy of the outcome distribution is replaced by
$H(R|B)$, the conditional von Neumann entropy of the post-measurement
state (after $R$ is measured) given $B$. This quantifies the uncertainty
about the outcome of a measurement of $R$ given access to $B$ (see the
Appendix). This relation is a strict generalization
of~\eqref{MaassenUffink} and features an additional term on the
right-hand-side. This term is a measure of how entangled the system
$A$ is with the observer's particle, $B$, expressed via the
conditional von Neumann entropy of the joint state, $\rho_{AB}$ of $A$
and $B$ before measurement, $H(A|B)$.  Note that this quantity can be
\emph{negative} for entangled states and in this case lowers the bound on
the sum of the uncertainties.  In particular, if $\rho_{AB}$ is
maximally entangled, $H(A|B)=-\log_2 d$, where $d$ is the dimension of
the system. Since $\log_2 1/c$ cannot be larger than $\log_2
d$, the RHS of~\eqref{Berta} cannot be greater than zero
for a maximally entangled state and, as mentioned previously, both $R$
and $S$ are perfectly predictable in such a case, for any observables
$R$ and $S$.  From a fundamental point of view, this highlights the
additional power an observer holding quantum information about the
system has compared to an observer holding classical information.

\begin{figure*}[t!]
\begin{center}
\includegraphics[width=1.8\columnwidth]{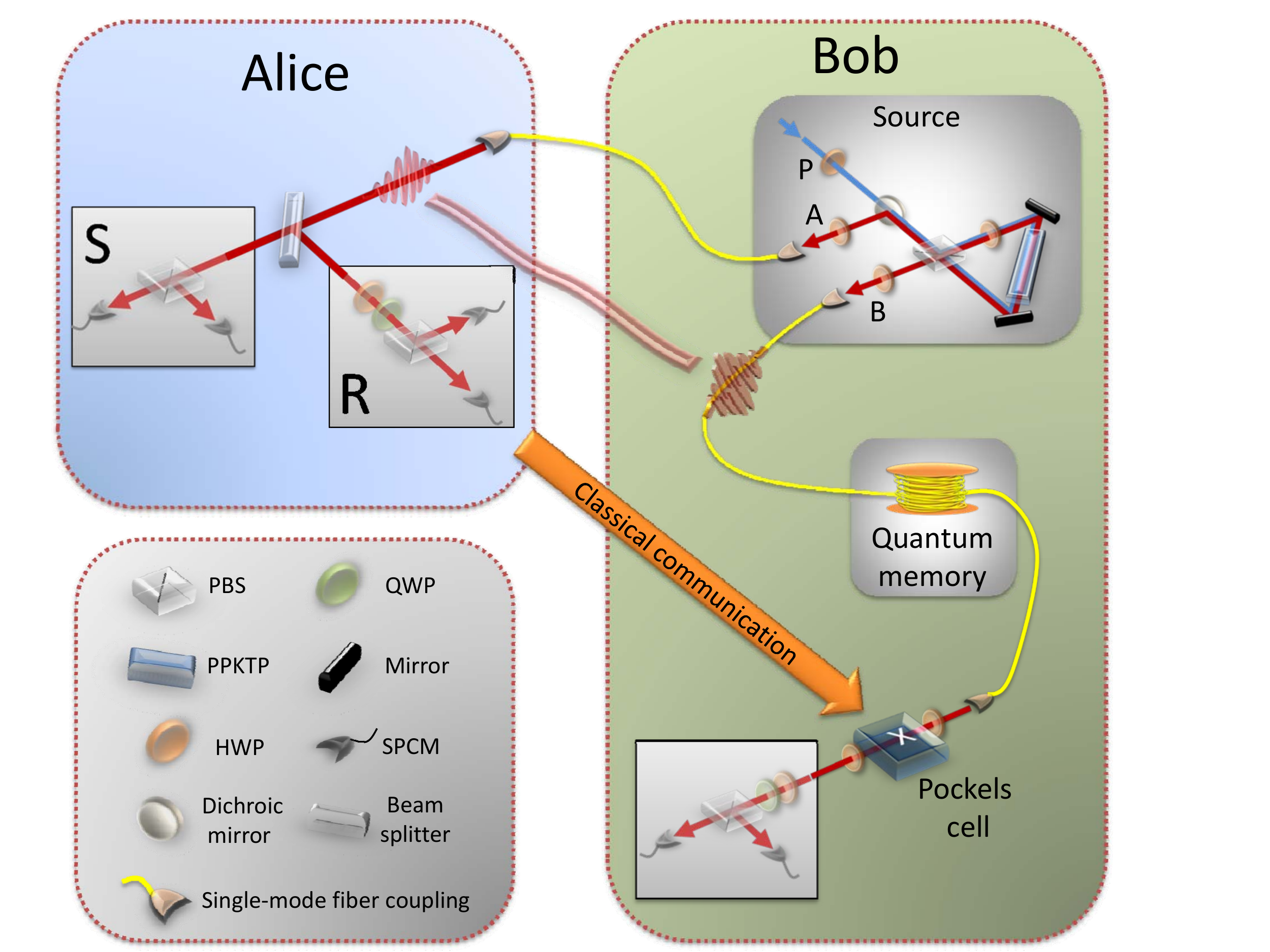}
\end{center}
\caption{Schematic of the experiment. Entangled photon pairs
  are created by pumping a periodically poled KTP crystal (PPKTP)
  inside a Sagnac interferometer and are subsequently
  fibre-coupled. Half-wave plates (HWP) $P$, $A$, and $B$ are used to
  prepare arbitrary polarization entangled states (see Appendix for
  details). One photon is then distributed to Alice who measures one
  of two observables, $R$ or $S$ (corresponding to different
  polarization bases). The observable is randomly selected by a 50/50
  beamsplitter (BS) which separates the two polarization analyzer
  modules. The choice of basis is classically transmitted back to Bob, who in
  the meantime delays his photon in a $50\m$ single-mode fibre. A fast Pockels cell (PC),
  which performs the identity when off or a bit flip operation
  ($\sigma_x$) when on, in combination with HWPs adapts Bob's
  measurement basis accordingly. The photons are detected using
  single-photon counting modules (SPCMs) and coincidence events
  between Alice's and Bob's detectors are recorded. Here we have depicted the
  experiment in the case that Bob measures his photon in the same
  basis as Alice.  In some runs of the experiment, we also perform
  tomography on Bob's photon (see main text). QWP, quarter-wave
  plate; PBS, polarizing beam splitter.}
\label{fig:1}
\end{figure*}

In order to clarify the sense in which an observer holding quantum
information can outperform one without, we follow Berta \emph{et
al.}~\cite{Berta2010} and consider uncertainty relations in the form
of a game between two parties, Alice and Bob: Bob creates a
quantum system and sends it to Alice.  He can prepare this system as
he likes and, in particular, it can be entangled with another particle
which he stores in a quantum memory (a device that maintains the
quantum coherence of its content). Alice then performs one of two
pre-agreed measurements, $R$ or $S$, chosen at random. She then
announces the chosen measurement, but not its outcome. Bob's aim is
to minimize his uncertainty (as quantified by the conditional von
Neumann entropy) about Alice's measurement outcome (see
Fig.~\ref{fig:1}).

In this work, we test the new inequality of Berta \emph{et al.}\
experimentally using entangled photon states and an optical delay
serving as a simple quantum memory. Entanglement allows us to
achieve lower uncertainties about both observables than would be
possible with only classical information over a wide range of
experimental settings. Our work addresses a cornerstone relation in
quantum mechanics and, to the best of our knowledge, is the first to
test one of its entropic versions. In the past, experiments have come close to the
original uncertainty limit~\cite{Elion1994,Nairz2002,LaHaye04,Schliesser2009}, but did not
involve entangled quantum systems. We also illustrate the practical
usefulness of the new inequality by applying it as an effective
entanglement witness.  

In our experiment, we use polarization-entangled photon pairs
generated by spontaneous parametric down-conversion (SPDC) and
polarization measurements on the individual photons to test the
inequality. Inferring entropic uncertainties from experimental data requires a high level of precision and control over the quantum system under consideration. Polarization-encoded photonic qubits offer this ability, making them a suitable testbed for the new uncertainty relation.

The schematics of the experiment and its connection to the uncertainty game are shown in Fig.~\ref{fig:1}. Our entangled photon pair source~\cite{Wong2006,Fedrizzi2007,Biggerstaff2009} can produce an entangled state of the form
\begin{equation}\label{SPDC}
  \ket{\Phi}_{AB}=\cos\zeta\ket{H_A H_B}+\sin\zeta\ket{V_A V_B},
\end{equation}
where $\ket{H}$ ($\ket{V}$) denotes a horizontally (vertically)
polarized photon and the subscripts label the spatial modes (Alice and
Bob, respectively). Control over the parameter $\zeta$ allows us to
change the amount of entanglement, characterized by the tangle
$\tau$~\cite{Hill1997}, between the two photons (see Appendix). We can
therefore study the inequality for a wide range of different
experimental settings.

In our experiment we realize Berta \emph{et al.}'s uncertainty game,
as shown in Fig.~\ref{fig:1}. The photon sent to Alice is entangled
with a second photon which is delayed by sending it through a $50\m$
single-mode fibre which acts as a quantum memory. This
gives Alice sufficient time to measure one of the two observables and
to communicate her measurement choice, but not the outcome, to Bob
before his photon emerges from the fibre (this is referred to as
\emph{feed-forward}).  On Bob's side, we either perform state
tomography, or have Bob measure his photon in the same basis as Alice.
In the latter case, a fast Pockels
cell~\cite{Prevedel2007,Biggerstaff2009} allows rapid switching
between measuring one of two pre-agreed observables. In total, the feed-forward
time is on the order of $\sim150\ns$.  More details regarding the
experiment can be found in Fig.~\ref{fig:1} and in the Appendix.
\begin{figure*}[p!]
\begin{center}
\includegraphics[width=1.7\columnwidth]{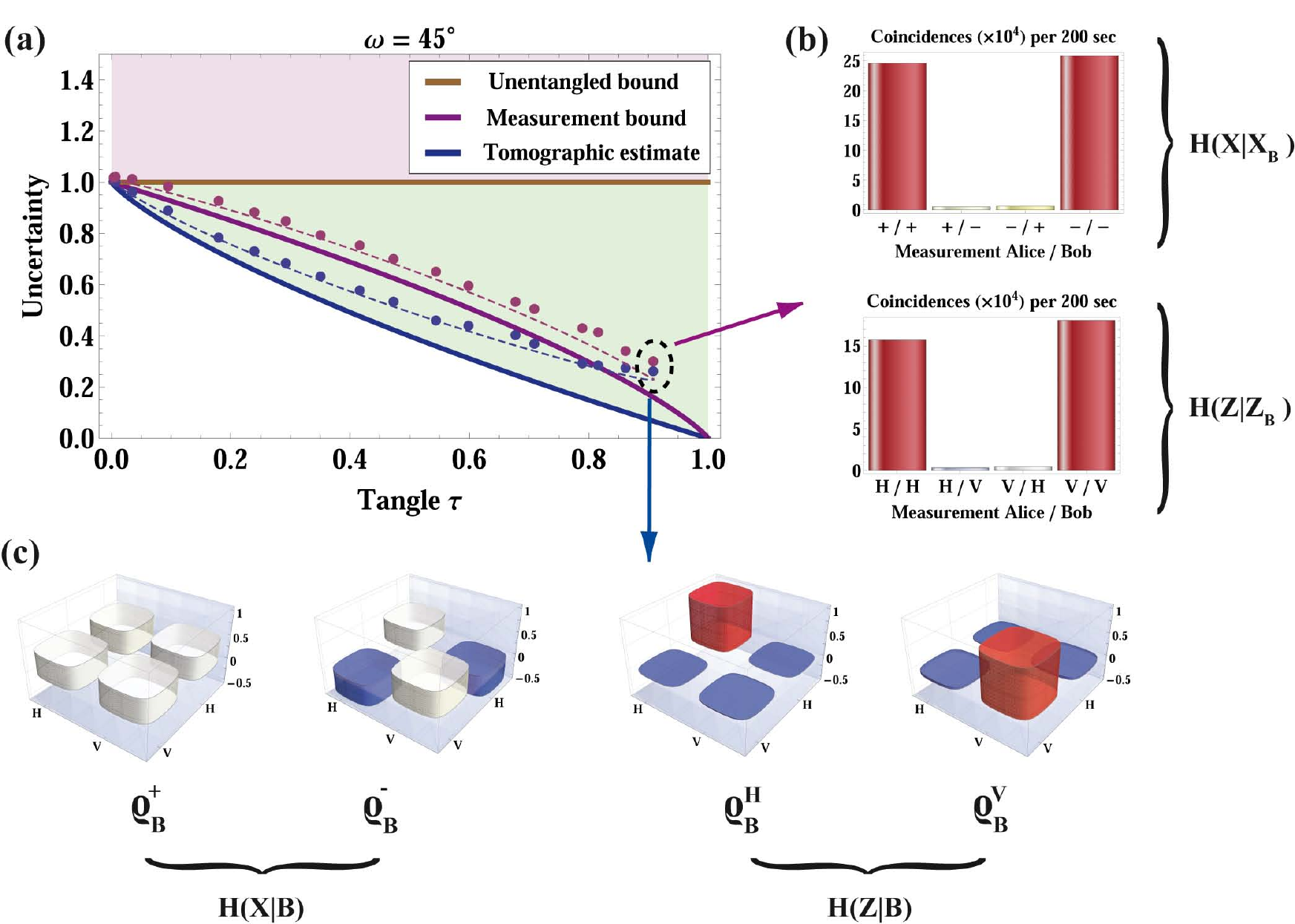}
\end{center}
\caption{Experimental results. In (a) we plot the left-hand
  side (LHS) of the new inequality for the case
  where $R=X$ and $S=Z$. In this case the relative angle, $\omega$,
  between the observables is $\omega=\pi/4$ ($45^{\circ}$). Varying HWP
  $S$ in the source allows us to plot the LHS for states with varying
  entanglement $\tau$. To calculate $H(X|B)+H(Z|B)$ we evaluate the
  entropies of the conditional single-qubit density matrices of Bob's
  qubit which we obtain through quantum state tomography (blue
  dots). Experimentally, this is achieved by setting the analyzers
  on Alice's side to perform measurements in the $\{\ket{H},\ket{V}\}$
  and $\{\ket{+},\ket{-}\}$ bases. If we also perform projective measurements on Bob's side we can obtain the entropies $H(X|X_B)+H(Z|Z_B)$ (purple dots)
  directly from the obtained coincidence count rates. Solid lines represent the theoretical bounds for the two entropy calculations, while the dashed lines indicate the simulated performance of the experiment. Note that for conjugate observables, the tomographic estimate coincides with equation~\eqref{Berta}. For the datapoints associated with the highest entanglement we show the corresponding coincidence count  rates in (b) and the reconstructed conditional density matrices in
  (c). Error bars ($\sim10^{-4}$) are too small to be seen. See
  Appendix for details.}
\label{fig:2}
\end{figure*}

\begin{figure*}[p!]
\begin{center}
\includegraphics[width=1.7\columnwidth]{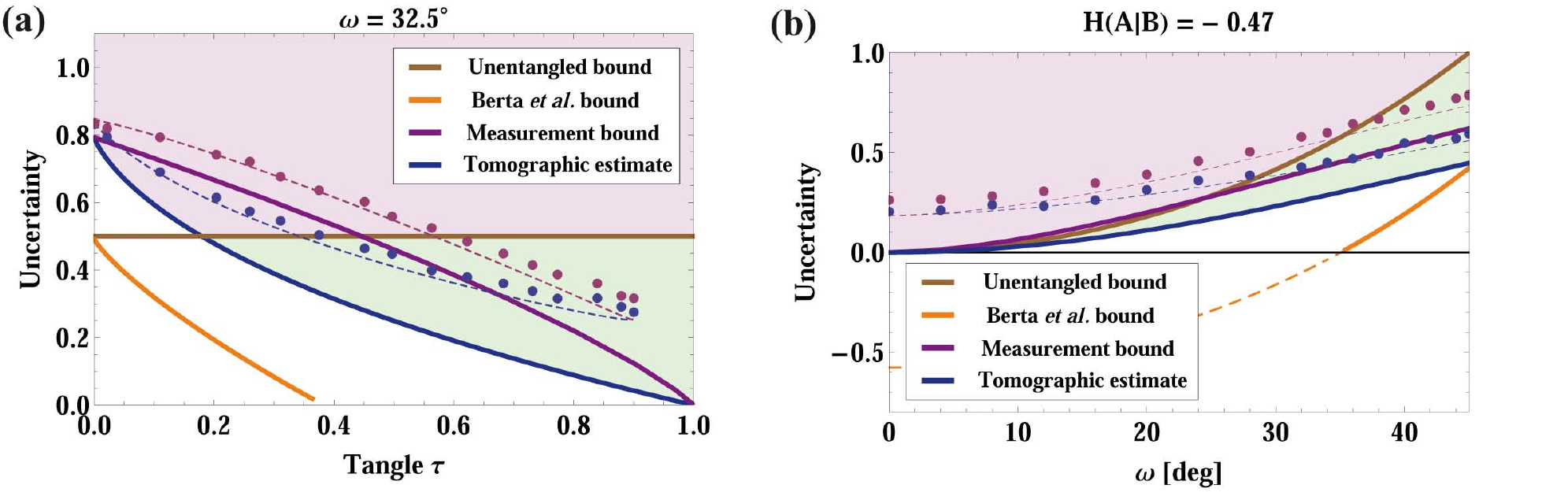}
\end{center}
\caption{Uncertainties for other experimental settings. In
  (a) we fix the relative angle between two measurement bases on
  Alice's photon at $\omega\approx0.57$ (32.5$^{\circ}$). This
  corresponds to the case where the bound in
  equation~\eqref{MaassenUffink} drops to 1/2. Note that here the new
  uncertainty relation~\eqref{Berta} is no longer tight. In (b) we
  chose a non-maximally entangled state (with $\tau\approx0.47$) and
  vary $\omega$, i.e. we fix $S=Z$ and vary $R(\omega)$ from $0$ to
  $45^{\circ}$.  We remark that the conditional entropies $H(R|B)$ and
  $H(S|B)$ cannot be negative and so, in the cases where the RHS
  of~\eqref{Berta} is negative, the bound $H(R|B)+H(S|B)\geq 0$ should
  be used instead.}
\label{fig:3}
\end{figure*}

The results of our experimental investigation are shown in
Figs.~\ref{fig:2} and~\ref{fig:3}.  The difference between the
original uncertainty principle, equation~\eqref{MaassenUffink}, and
Berta \emph{et al.}'s result, equation~\eqref{Berta}, is most apparent
for the case of maximal entanglement and
conjugate observables, i.e.\ for $R=X$ and $S=Z$.  In this scenario, Bob can predict the
outcome of Alice's measurement perfectly, i.e.\ $H(X|B)+H(Z|B)=0$,
which would be impossible if Bob did not have a quantum memory (the
RHS of equation~\eqref{MaassenUffink} has $\log_{2}1/c=1$ for these
observables).  In fact, for any finite entanglement between Alice's
particle and Bob's quantum memory we expect to find lower
uncertainties than in the case of no entanglement.  This trend is
clearly observed in Fig.~\ref{fig:2} where we vary the entanglement
(characterized by the tangle $\tau$) for the case of conjugate
observables. This shows that entanglement allows Bob to
predict both observables more precisely than without. We also use two
different approaches to estimate the LHS of
equation~\eqref{Berta}. The first is a direct determination of
$H(X|B)+H(Z|B)$ (the blue, solid line in~\ref{fig:2}(a)) which
requires calculation of the reduced density matrix of Bob's photon for
each of the alternative measurement choices and outcomes, which can in
turn be obtained through quantum state tomography~\cite{James2001}.
Alternatively, we can bound the entropies by also performing a
projective measurement on Bob's photon, which allows us to estimate
$H(X|X_B)+H(Z|Z_B)$ (where $X_B$ and $Z_B$ are the observables
measured by Bob). Since $H(X|B)\leq H(X|X_B)$, this technique will in
general only provide an upper bound on $H(X|B)+H(Z|B)$ and therefore
yield a weaker inequality.  Its advantage is that it can
be estimated with a straightforward experimental test without
tomography. In Fig.~\ref{fig:2} we show the results of both experimental approaches and we outline the details of the entropy calculations in the Appendix.

Furthermore we investigate the new uncertainty relation for other
choices of observables. Choosing non-conjugate observables lowers the
RHS of both inequalities~\eqref{MaassenUffink} and~\eqref{Berta}. In
Fig.~\ref{fig:3}(a) we chose the relative angle between the
observables, $\omega$, as $\omega\approx0.57$ ($32.5^{\circ}$) which
is where the unentangled bound decreases to $\log_2 1/c=1/2$.
Inequality~\eqref{Berta} is not tight in this case, i.e.\ there is no
state for which equation~\eqref{Berta} is satisfied with
equality. This is seen in Fig.~\ref{fig:3}, where the tomographic
estimate no longer coincides with the Berta \emph{et al.}\ bound (as
it did for the case of conjugate observables). This scenario places
more stringent requirements on the quality of the experiment in order
to show that the entanglement allows for lower
uncertainties. Nevertheless we find lower entropies than predicted by
inequality~\eqref{MaassenUffink} for sufficiently large
entanglement. Discrepancies from the ideal, theoretical bound are
mainly due to the imperfect entanglement between the
photons. Simulations of the experiment, based on the measured
fidelities ($F\approx0.97$) of our entangled photon pair source and assuming white
noise as the dominant source of imperfection, confirm this fact (see dashed lines in Figs.~\ref{fig:2} and~\ref{fig:3}).

In Fig.~\ref{fig:3}(b), we investigate the new uncertainty relation
for a fixed partially entangled state, varying the complementarity of
the observables, $c(\omega)$.  Again we find good agreement and
uncertainties consistent with the new inequality, thereby providing
strong evidence for the validity of the new uncertainty principle in
practice. Further discussion on the optimization of the entangled states and measurements required to most stringently test the uncertainty principle are described in the Appendix.


We now discuss our experiment and the new inequality in the context of
the proposed application as an entanglement witness.
Uncertainty relations have been used in the past to derive entanglement witnesses~\cite{Hofmann2003,Guehne2004}. The idea in our case is to use equation~\eqref{Berta} to bound $H(A|B)$. Whenever $H(R|B)+H(S|B)<\log_2 1/c$,
we can conclude that $H(A|B)<0$ which is a certificate that
$\rho_{AB}$ is entangled. This is readily observed in
Figs.~\ref{fig:2} and~\ref{fig:3}: any datapoint below the unentangled
bound indicates the presence of entanglement.

The best entanglement witness is for the case of complementary observables.  As can be seen in Fig~\ref{fig:2}(a), the quality of the witness depends on the technique used.  In the case that Bob measures, our experiment detects entanglement for $\tau\geq0.06\pm0.01$, which is higher than the analogous bound obtained with tomography, $\tau\geq0.007\pm0.003$.  However, using tomography requires estimation of more parameters (16 vs 8).  An even simpler bound can be obtained using only 2 parameters, which we find to be only slightly weaker as a witness: it detects entanglement for $\tau\geq0.13\pm0.02$ (see Appendix Fig.~\ref{fig:supp}).

Significantly, the 2 parameters needed for the simpler bound can be obtained using local measurements, making it a very simple entanglement witness.  For single qubits, the merits of this are minor when compared to full tomography.  However, more generally the separation between the number of parameters scales like the square of the dimension of the system, making the tomographic estimate infeasible for systems comprising more than a few qubits.
We further remark that,
although one parameter witnesses exist, these usually require
measurement of a joint operator, which can be difficult to implement.
In practice these witness operators are often first decomposed into locally measurable parts~\cite{Terhal2002,Guehne2002,Guehne2009}.

Until recently, all known bounds on the uncertainty an observer can
have about the outcomes of measurements on a system applied only to
observers holding classical information about the system. Berta
\emph{et al.}\ have since overcome this limitation, deriving a
stronger uncertainty relation which applies when one observer holds
quantum information about another system in a quantum memory. In this work, we
give the first experimental investigation of this strengthened
relation.  We demonstrate that entangling the system with a particle
in a quantum memory does indeed lead to lower bounds on the
uncertainty than is possible without. Our results also quantitatively
illustrate the theoretical behaviour of the new uncertainty relation,
with discrepancies explained from the measured quality of our source.
Future improvements in both photon sources and detectors will allow more
precise tests of its bounds.  Additionally, since we achieve lower
uncertainties than would be possible without entanglement, our
experimental setup acts as an effective entanglement witness, and
succeeds as such over a wide range of entanglement.

We thank M. Piani for valuable discussions and the Ontario Ministry of
Research and Innovation ERA, QuantumWorks, NSERC, OCE, Industry Canada
and CFI for financial support. R.P. acknowledges support by MRI and
the Austrian Science Fund (FWF).

\bibliographystyle{apsrev}

\appendix

\section{Appendix}

\subsection{Entropy inference}
In this section we give an account of how the quantities appearing in
equation~\eqref{Berta} can be inferred from the data obtained in the
experiment.  We begin with the mathematical definition of the relevant
quantities.  For a density matrix $\rho_{AB}$, the von Neumann entropy
is defined by $H(AB):=-\tr(\rho_{AB}\log_2\rho_{AB})$, which is
conveniently calculated from the eigenvalues, $\{\lambda_i\}$, of
$\rho_{AB}$ by
$H(AB)=H(\{\lambda_i\}):=-\sum_i\lambda_i\log_2\lambda_i$.  For a
state $\rho_{AB}$, the conditional entropy of $A$ given $B$ is defined
as $H(A|B):=H(AB)-H(B)$, where $H(B)$ is the von Neumann entropy of
the reduced density operator, $\rho_B:=\tr_A\rho_{AB}$.  The quantity
$H(R|B)$ is the conditional von Neumann entropy of the state
\begin{equation}\label{condentropy}
\rho_{RB}:=\sum_r\left(\ketbra{\Psi_r}{\Psi_r}\otimes\id\right)\rho_{AB}\left(\ketbra{\Psi_r}{\Psi_r}\otimes\id\right),
\end{equation}
where $R$ corresponds to a measurement on the $A$ system in the
orthonormal basis defined by $\{\ket{\Psi_r}\}$ (this state is to be
interpreted as the post-measurement state after $R$ is measured).  It
will be convenient to write this in the following form:
\begin{equation}\label{condentropy}
\rho_{RB}=\sum_rp_r\ket{\Psi_r}\bra{\Psi_r}\otimes\rho_B^{r},
\end{equation}
where $p_r$ is the probability of obtaining outcome $r$ when $R$ is
measured, and $\rho_B^r$ is the state of the $B$ system when $r$
occurs. The relevant entropy can then be calculated using~\cite{Nielsen2000}
\begin{eqnarray*}
H(R|B)=H\left(\{p_r\}\right)+\sum_rp_rH\left(\rho_{B}^r\right)-H\left(\sum_rp_r\rho_{B}^r\right).
\end{eqnarray*}
The entropy $H(S|B)$ can be analogously defined.

The density operators $\{\rho_B^r\}_r$ are obtained by performing
tomography on the state of the $B$ system conditioned on a particular
outcome (see the next section for details).  This generates the
\emph{tomographic estimate} of the uncertainty.

Alternatively, we can estimate $H(R|B)$ by performing a measurement on
$B$ in a basis which we denote by $R_B$ with outcome $r_B$.  Since
$H(R|B)\leq H(R|R_B)$, i.e.\ measurements cannot decrease the entropy,
we in general obtain a higher uncertainty. The entropy $H(R|R_B)$ can be calculated from the resulting joint probability distribution of both measurements, $P(r,r_B)$, via $H(R|R_B)=H(\{P(r,r_B)\})-H(\{P(r_B)\})$.  This gives rise to the
\emph{measurement bound} on the uncertainty.

We also calculate the entropy using the bound $H(R|R_B)\leq-q_R\log_2
q_R-(1-q_R)\log_2(1-q_R)$ (which comes from Fano's inequality), where
$q_R$ is the probability that $r\neq r_B$.  This gives rise to the
\emph{Fano bound} on the uncertainty. See the below for more information.

In the experiment we investigate the uncertainties for two-qubit
states with Schmidt coefficients $\cos\zeta$ and $\sin\zeta$.  Such
states have conditional von Neumann entropy
$H(A|B)=-H(\{\cos^2\zeta,\sin^2\zeta\})$ and tangle $\tau=\sin^2 2\zeta$.

\subsection{Experiment}
In our experiment, we generate the entangled photons pairs using type-II spontaneous parametric down-conversion (SPDC). A 0.7$\mW$ diode laser at 404$\nm$ pumps a 25$\mm$  periodically-poled KTiOPO$_4$ (PPKTP) crystal in a Sagnac configuration, emitting entangled photons which are subsequently single-mode fibre-coupled after 3$\nm$ bandpass interference filters (IF) \cite{Wong2006,Fedrizzi2007,Biggerstaff2009}. Typically we observe a coincidence rate of 15$\kHz$ directly at the source. A half-wave plate (HWP) $P$ before the Sagnac interferometer controls the pump polarization and therefore allows us to precisely control $\zeta$ in equation~\eqref{SPDC} and hence the entanglement of the generated state. Additional HWPs at the outputs of the fibres rotate the entangled state into the desired Schmidt basis $\ket{\theta}$ (see below). Photons are detected by single-photon counting modules (SPCM) and their frequencies are recorded using a multichannel logic with a coincidence window of 3$\ns$.

On Alice's side, two polarization analyzer modules, each consisting of
a PBS preceded by a QWP and HWP, are separated by a 50/50
beamsplitter. One of them is set to measure in the
$\{\ket{H},\ket{V}\}$ basis while the other can be set to measure at
some chosen angle in the X-Z plane,
i.e.\ $\{\ket{\omega},\ket{\omega^{\bot}}\}$ where the $\omega$ in
$\ket{\omega}=\cos\omega\ket{H}+\sin\omega\ket{V}$ is the angle of the
linear polarization.

The other down-converted photon is meanwhile delayed in a 50$\m$ single-mode optical fibre spool, which is long enough to execute the measurements on Alice's side and communicate (feed-forward) her chosen basis to Bob. Depending on the basis, Bob switches between two analyzer bases, $R$ and $S$. This is achieved by using a fast RbTiOPO$_4$ (RTP) Pockels cells (PC), aligned so as to perform a $\sigma_x$ (X) operation~\cite{Prevedel2007,Biggerstaff2009}. HWPs before and after the PC allow to adapt the switchable analyzer bases. Therefore, after passing the PC, Bob's photon is effectively measured in the $\{\ket{\omega},\ket{\omega^{\bot}}\}$ ($\{\ket{H},\ket{V}\}$) basis when the PC is on (off).

The experiment itself is performed as follows. At the start of each
run, quantum state tomography~\cite{James2001,Langford2005} is
performed on the entangled photon pair. We record coincidences between
Alice's reflected arm of the BS and Bob's polarization analyzer
following the switched off PCs. Coincidence measurements were
integrated over $5\s$ for a tomographically over-complete set of
measurements, comprising all combinations of the six eigenstates of
$X$, $Y$, and $Z$ on Alice's and Bob's qubit, respectively. Using an
iterative technique~\cite{Jezek2003} we reconstruct the density matrix
$\rho_{\text{AB}}^{\text{exp}}$, from which we infer the tangle $\tau$
of our state. We then set the analyzers on Alice's side to the
$\{\ket{H},\ket{V}\}$ ($\{\ket{\omega},\ket{\omega^{\bot}}\}$) basis
in the transmitted (reflected) arm of the BS and perform conditional
single-qubit tomography on Bob's photon, from which we calculate
$H(R(\omega)|B)+H(Z|B)$. Finally, Bob's analyzer is set to the
$\{\ket{H},\ket{V}\}$ basis which allows us to calculate $H(Z|Z_B)$
and $H(R(\omega)|R_B(\omega))$ directly from the coincidence
counts. Stepwise repetition of this procedure for varying $\zeta$ or
$\omega$ leads to the data presented in Figs.~\ref{fig:2}
and~\ref{fig:3}.

\subsection{State and measurement optimization}
Our aim is to rigorously test the validity of the new uncertainty
relation~\eqref{Berta} in an experimental setting.  However, it is not
the case that for all pairs of measurements there exists a state which
saturates the bound. Likewise, it is not the case that for all states
the bound can be saturated by some pair of measurements. Hence, in
order to probe the bound, we try to observe the minimum uncertainties
possible (i.e.\ the minimum left-hand side attainable). In this
section, we show how to achieve this minimum.

We start by considering the new uncertainty relation in the case where
the state of the system and memory is a pure two-qubit state.  Without
loss of generality, we can assume $S$ is a measurement in the
$\{\ket{H},\ket{V}\}$ basis and $R$ is a measurement in the $\{\cos
\omega\ket{H}+\sin \omega\ket{V},-\sin
\omega\ket{H}+\cos\omega\ket{V}\}$ basis, which we denote
$\{\ket{\omega},\ket{\omega^{\bot}}\}$. The pure two-qubit state
$\ket{\Phi_{AB}}$ on which we apply the relation can be written in its
Schmidt basis
\begin{equation*}\label{schmidtstate}
\ket{\Phi_{AB}}=\cos\zeta\ket{\theta}_A\ot\ket{\theta}_B+\sin\zeta\ket{\theta^{\bot}}_A\ot\ket{\theta^{\bot}}_B,
\end{equation*}
where $\ket{\theta}$ and $\ket{\theta^\bot}$ are orthogonal states,
which we generically write as $\ket{\theta}=\cos\theta\ket{H}+e^{i\varphi}\sin\theta\ket{V}$ and $\ket{\theta^\bot}$ being the orthogonal state. Using the binary entropy function $h_2(p):=-p\log_2 p-(1-p)\log_2(1-p)$, we can write out the terms in equation~\eqref{Berta} as
\begin{eqnarray*}
H(A|B)&=&-h_2(\cos^2\zeta)\\
c&=&\max\left(\cos^2\omega,\,\sin^2\omega\right)\\
H(R|B)&=&h_2\Big{(}\frac{1}{2}(1-\cos(2\zeta)(\cos(2\omega)\cos(2\theta)\\
& &+\sin(2\omega)\sin(2\theta)\cos\varphi))\Big{)}-h_2(\cos^2\zeta)\\
H(S|B)&=&h_2\left(\frac{1}{2}(1-\cos(2\zeta)\cos(2\theta))\right)-h_2(\cos^2\zeta).
\end{eqnarray*}
For fixed entanglement $H(A|B)$, i.e.\ fixed $\zeta$, and a given
complementarity between the observables, i.e.\ fixed $\omega$, we want
to find the corresponding entangled state $\ket{\Phi_{AB}}$ that
achieves the minimum uncertainty, so that we can get closest to the
bound given by equation~\eqref{Berta}.

\subsubsection{Conjugate observables}
This is the case $R=X$, i.e.\ $\omega=\frac{\pi}{4}$ so that $H(R|B)$
reduces to
$$H(R|B)=h_2\left(\frac{1}{2}(1-\cos(2\zeta)\sin(2\theta)\cos\varphi)\right)-h_2(\cos^2\zeta).$$
The minimum over $\varphi$ is then for $\varphi=0$.  We then have
\begin{eqnarray*}
&&H(R|B)+H(S|B)=h_2\left(\frac{1}{2}(1-\cos(2\zeta)\sin(2\theta))\right)\nonumber\\
&&\quad+h_2\left(\frac{1}{2}(1-\cos(2\zeta)\cos(2\theta))\right)-2h_2(\cos^2\zeta).
\end{eqnarray*}
From the form of $h_2(p)$, it is clear that the minimum over $\theta$
occurs for $\theta=0$ or $\theta=\frac{\pi}{4}$ and has the value
$1-h_2(\cos^2\zeta)$.  In other words, for the case $R=X$, $S=Z$, and
fixed $H(A|B)$, to minimize the uncertainty, the best choice of state
is $\ket{\Phi}=\cos\zeta\ket{HH}+\sin\zeta\ket{VV}$.  The parameter
$\zeta$ is related to the tangle $\tau$ through the relation
$\tau=\sin^{2}2\zeta$ and can be conveniently set by HWP $P$ in our
photon pair source. Note that, in this case, the bound given by
equation~\eqref{Berta} is achievable.  This is the blue, solid line
(tomographic bound) in Fig.~\ref{fig:2}(a).

\subsubsection{General observables}
\begin{figure}[b]
\includegraphics[width=1\columnwidth]{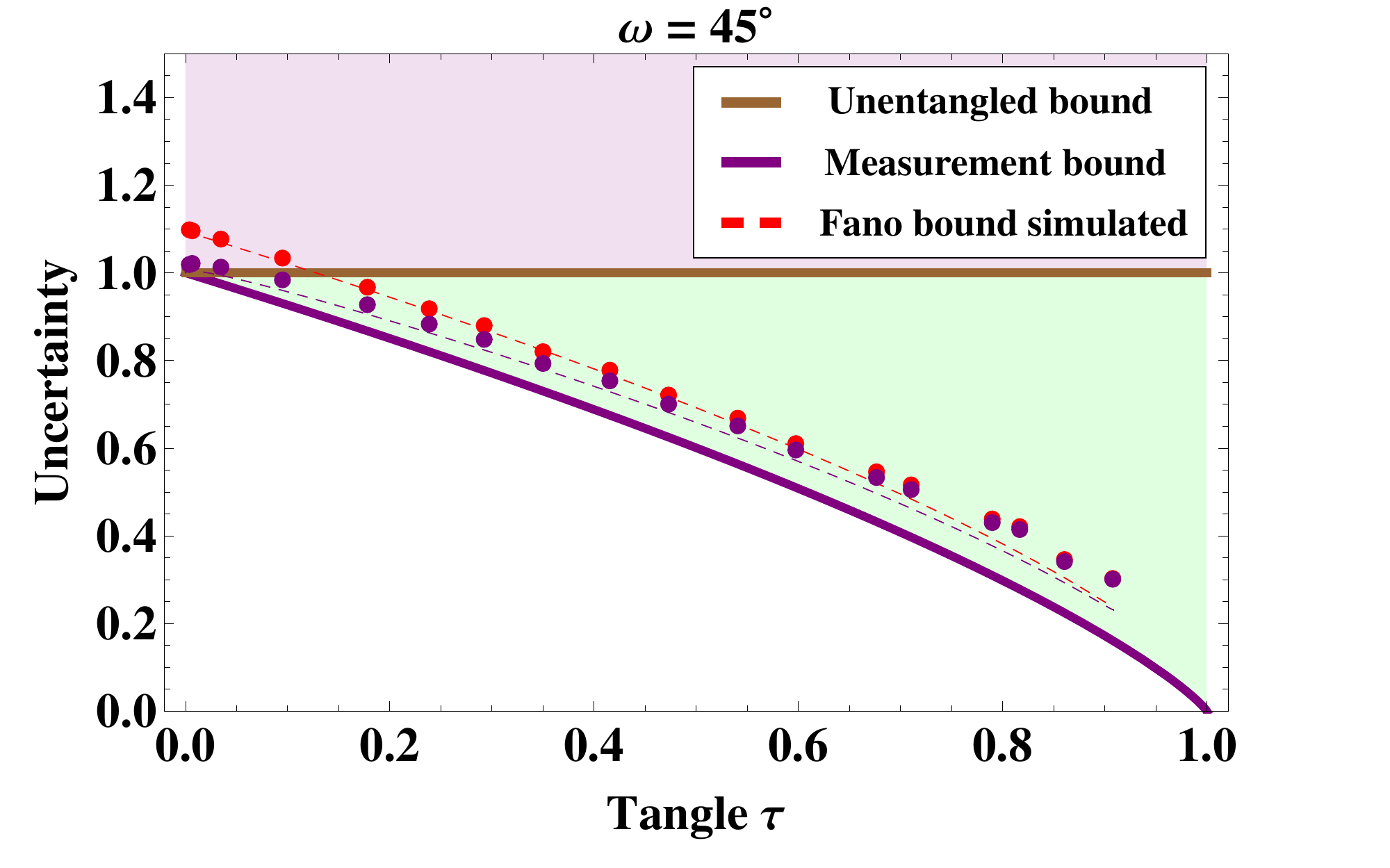}
\caption{Comparison of the measurement bound and the Fano bound.
  Here, we compare the two bounds in the case of complementary
  observables (c.f.\ Fig.~\ref{fig:2}(a)). Theoretically, the bounds
  coincide for this case (purple line). However, in the experiment, we
  find a slight difference and a worse estimate for the Fano bound
  (red dots) when compared to our standard method (purple
  dots). Simulations of the experiment (dashed lines) are in good
  agreement with the data.}
\label{fig:supp}
\end{figure}

In the general case of arbitrary $R$, the optimal $\varphi$ can be $0$ or
$\pi$ depending on the other parameters.  However, we can always take
$\varphi=0$ and note that the minimum over $\theta$ accounts for the two
possibilities (taking $\varphi$ from $0$ to $\pi$ is equivalent to taking
$\theta$ to $\pi-\theta$, in terms of the entropies). The task is
then to minimize
\begin{align*}
&h_2\left(\frac{1}{2}\left(1-\cos(2\zeta)(\cos(2\omega)\cos(2\theta)+\sin(2\omega)\sin(2\theta))\right)\right)\\
&+h_2\left(\frac{1}{2}(1-\cos(2\zeta)\cos(2\theta))\right)-2h_2(\cos^2\zeta)
\end{align*}
on $\theta$ for fixed $\zeta$ and $\omega$.  For a wide range of
parameters, the minimum occurs for $\theta=\omega/2$, so that the best
entangled state is aligned ``in between'' the $Z$ axis and the axis of
$R$.  However, when the measurements are close to complementary (i.e.\
$\omega\approx\frac{\pi}{4}$), the minimum can occur at $\theta=0$ (as
in the case of perfectly complementary observables).

The optimal measurement of Bob is not necessarily the same as that of
Alice in the general case. In general, the best measurement setting on
Bob's side can be found by numerical optimization. However, in the
cases investigated here, choosing the same measurement on both sides
provides an entropy close to that of the optimal with the difference
being insignificant when compared to experimental errors.

\end{document}